# BlueBird 6 is Fainter than Block 1 Satellites

Anthony Mallama [1] and Richard E. Cole [1]

[1] IAU Centre for the Protection of the Dark and Quiet Sky,
IAU-UAI Headquarters, 98-bis Blvd Arago, 75014,
Paris, France

Correspondence: anthony.mallama@gmail.com

The mean apparent magnitude for BlueBird 6, the first Block 2 satellite of the BlueBird constellation, is 3.85 +/- 0.10 while that for Block 1 spacecraft is 3.30 +/- 0.07. So, BlueBird 6 is 0.55 magnitudes fainter. Likewise, the means of apparent magnitudes adjusted to a uniform distance of 1,000 km are 4.32 +/- 0.08 and 3.77 +/- 0.06, respectively. This difference is also 0.55 magnitude.

The dimming is unexpected because the Block 2 satellites are ~3.5 times as large as Block 1. BlueBird 6 is only 17% as bright as Block 1 spacecraft per unit surface area. Possible explanations are discussed.

All BlueBird satellites exceed the brightness limits recommended by the International Astronomical Union. However, they will be fewer in number than other satellite constellations.

## 1. Introduction

Bright satellites interfere with astronomical observations (Barentine et al. 2023) and spoil the aesthetic beauty of the night sky (Mallama and Young 2021). This concern has worsened with the launch of large satellite constellations.

Brightness statistics for BlueBird (BB) Block 1 satellites were reported by Cole et al (2025) and Mallama and Cole (2025). This paper reports on the first of the Block 2 spacecraft which are ~3.5 times as large.

Section 2 provides general information about the BB satellites, including their sizes and orbits. Section 3 describes the observations analyzed in this study. Section 4 characterizes the brightness of BB 6 and Block 1 spacecraft. Section 5 discusses brightness mitigation of BB spacecraft. Section 6 addresses their impact on optical astronomy and Section 7 summarizes our results.

## 2. Background

The BB constellation of direct-to-cell spacecraft is operated by AST SpaceMobile. Their prototype satellite BlueWalker 3 (BW 3) was observed extensively by Nandakumar et al (2023) and by Mallama et al (2024). The spacecraft was folded into a compact object during launch and then deployed into a large flat antenna panel on orbit.

Both Nandakumar et al (2023) and Mallama et al (2024) found that BW 3 was very bright and that it was sometimes among the most luminous objects in the sky. The brilliancy of BW 3 prompted a strong response[1] from the International Astronomical Union.

Five BB Block 1 satellites were launched on 2024 December 9 into 500 km altitude orbits with $53^{o}$ inclinations. Their flat panels have the same 64 $m^2$ area when deployed as BW 3 according to AST[2].

The brightness of Block 1 satellites was characterized by Cole et al (2025) and by Miller et al (2026). These studies reached similar conclusions about their luminosity. Cole et al also developed a physical model for brightness to explain how Block 1 differs from BW 3. Specifically, their antenna elements are different from BW 3 and they reflect sunlight in a unique way as discussed in Section 5.

The first Block 2 satellite, BB 6, was launched on 2025 December 24 into an orbit that is nearly identical to the Block 1 spacecraft. Several more Block 2 satellites were launched recently. However, there are too few observations of the later ones to derive robust brightness statistics and they are not part of this study.

The flat panels of Block 2 satellites have an area of 223 $m^2$, according to AST[3], which is ~3.5 times that of Block 1 spacecraft. So, the Block 2 satellites were expected to be ~1.4 magnitudes brighter based on the size ratio.

## 3. Observations

Most of the data for this study was recorded by experienced visual observers using binoculars or their unaided eyes. Magnitudes were determined by comparing the spacecraft to nearby reference stars. The angular proximity between satellites and stellar objects accounts for variations in sky transparency and sky brightness. Mallama (2022) describes this method in more detail. Visual magnitudes were obtained by the authors as well as B. Young, J. Barentine and J. Respler.

A smaller number of electronic measurements were recorded by the MMT9 robotic observatory in Russia (Karpov et al. 2015 and Beskin et al. 2017). The magnitudes are within 0.1 of the Johnson V-band as discussed by Mallama (2021). Data was recorded at a frequency of 10 Hz which we averaged into 5 second means. Electronic and visual data are combined in the analyses reported below.

## 4. Brightness characterization

The mean, *M*, apparent magnitude for BB 6 after deployment is 3.85, with a standard deviation, *SD*, of 1.55 and a standard error of the mean, *SEM*, of 0.10, as listed in Table 1. Magnitudes standardized to a distance of 1000-km have M, SD and SEM values of 4.32, 1.31 and 0.08, respectively. The final column of the table indicates that there are 258 observations of BB 6.

The magnitude statistics for Block 1 satellites are also listed in Table 1. These values are slightly different than those reported by Cole el al (2025) because we have added more observations to our database since that study was published.

Comparison of the mean values in Table 1 reveals that BB 6 is 0.55 magnitudes fainter for both apparent and 1000-km measures.

[1] https://cps.iau.org/news/iau-cps-statement-on-bluewalker-3-global-astronomy-community-troubled-by-unprecedented-brightness-and-use-of-terrestrial-frequencies-from-space-of-recently-launched-bluewalker-3-satellite/
[2] https://ast-science.com/bluebird-1-5/
[3] https://ast-science.com/next-gen-bluebird/

Table 1. BB Magnitude Statistics

| | Apparent magnitude | | | 1000-km magnitude | | | |
|---|---|---|---|---|---|---|---|
| | **Mean** | **SD** | **SEM** | **Mean** | **SD** | **SEM** | **Obs** |
| **BB 6** | **3.85** | **1.55** | **0.10** | **4.32** | **1.31** | **0.08** | **258** |
| **BB 1-5** | **3.30** | **1.70** | **0.07** | **3.77** | **1.45** | **0.06** | **546** |
| **Diff.** | **0.55** | | | **0.55** | | | |

The distributions of apparent magnitudes for BB 6 and BB 1-5 are compared in Figure 1, and the distributions of 1000-km magnitudes are shown in Figure 2.

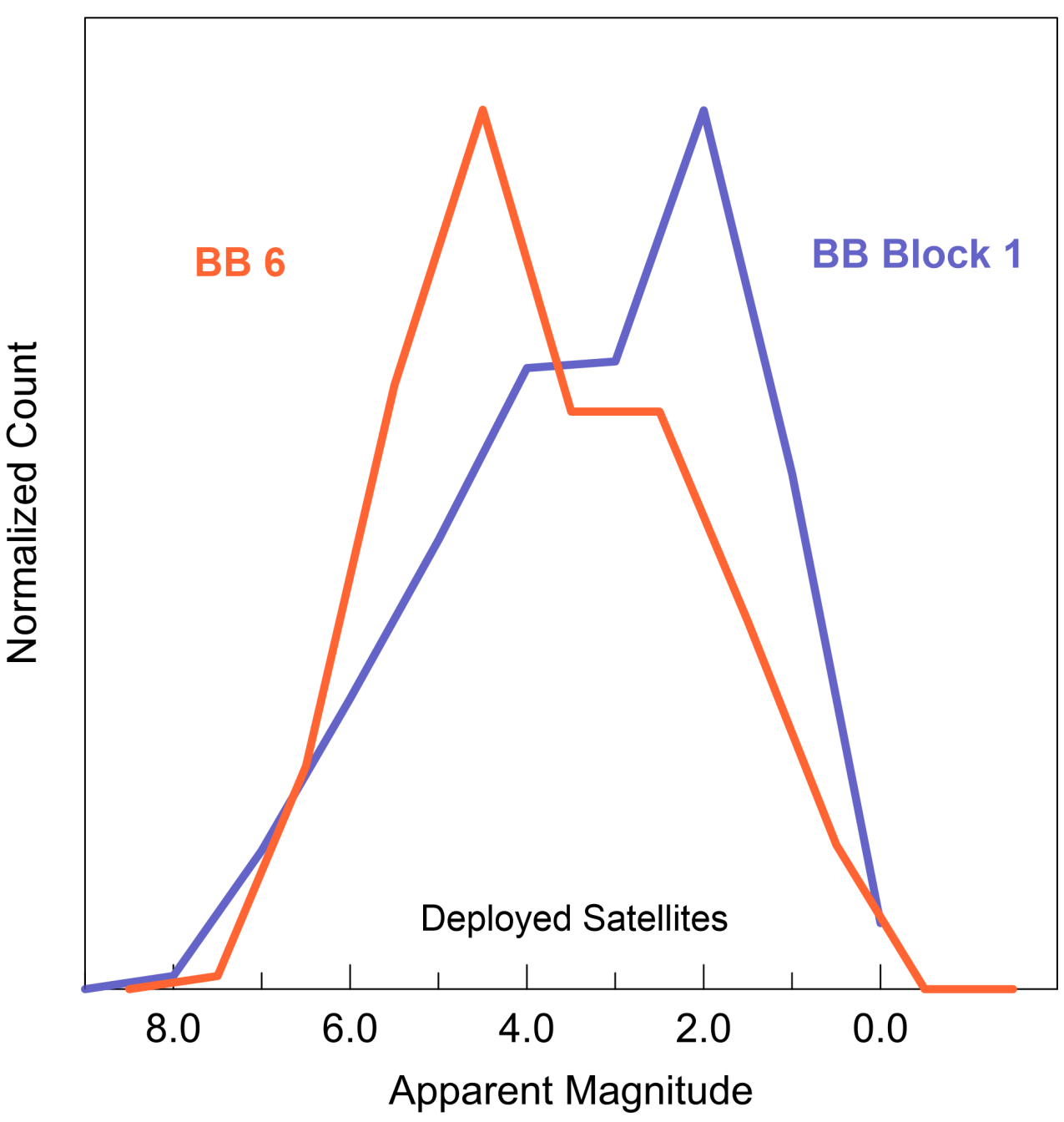

*Figure 1. The distribution of extinction corrected apparent magnitudes for BB 6 and Block 1 satellites. The peak of the distribution for BB 6 is fainter than that for Block 1.*

Besides magnitude distributions, other useful brightness functions include magnitudes versus (a) the satellite's height above the horizon, (b) the solar phase angle and (c) beta, the angle of the direction of the Sun to the orbital plane. These can aid in predicting a satellite's luminosity in the sky. The functions are also particular to different satellite constellations as well as different models such as BB Blocks 1 and 2.

Apparent magnitude is plotted versus elevation in Figure 3. BB satellites brighten as they reach greater elevations like most spacecraft. In this case, the brightening between 20° and 80° is about 3 magnitudes. One reason for this increased luminosity is that elevation is a proxy for distance and satellites overhead are closer. Also notice that the best linear fit to BB 6 magnitudes is fainter than that for BB 1 - 5 throughout the observed span of elevations.

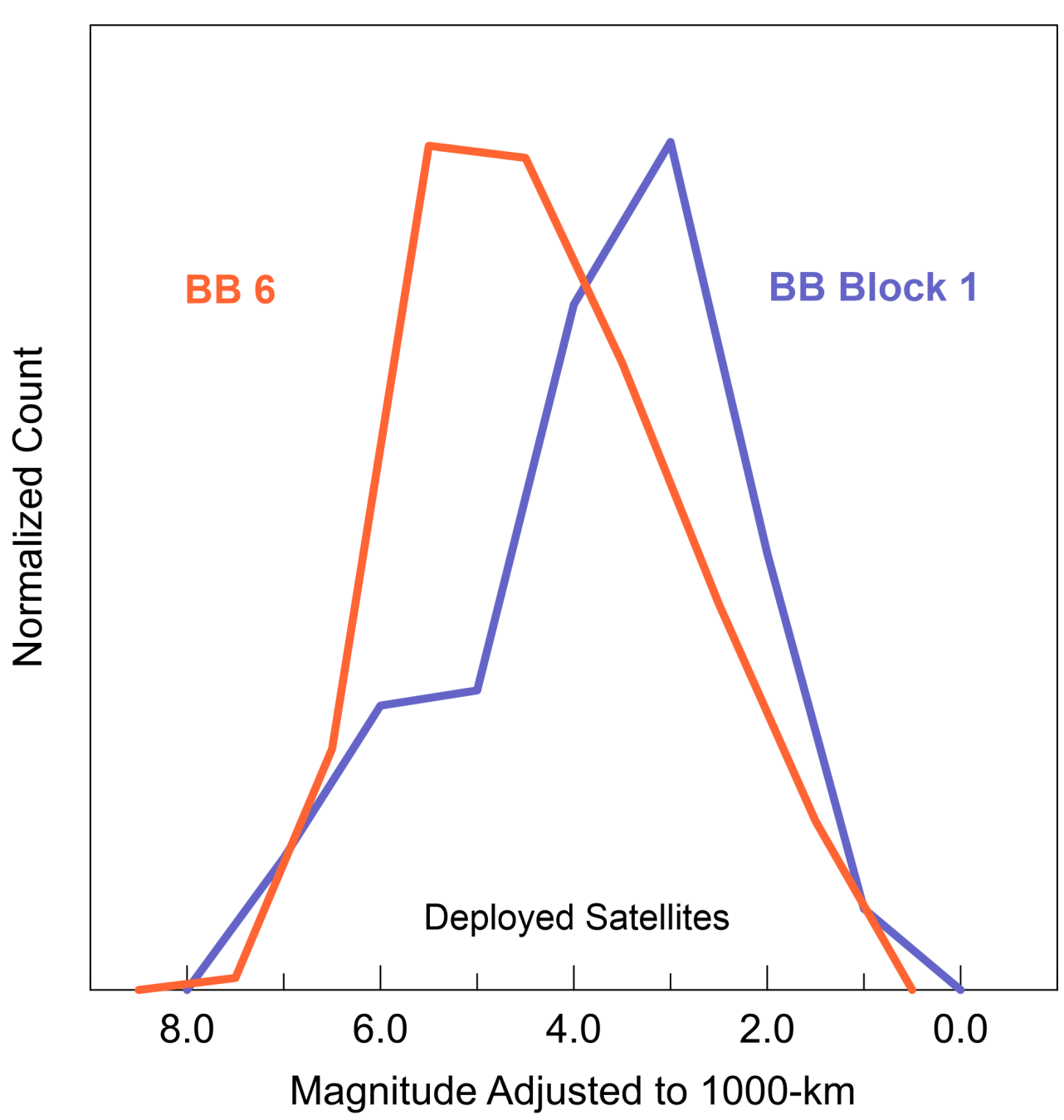

*Figure 2. The distribution of magnitudes adjusted to a standard distance of 1000-km. As with apparent magnitudes, the peak of the distribution for BB 6 is fainter than for Block 1.*

Phase angle is the arc measured at a satellite between vectors to the Sun and the observer. Apparent magnitudes are adjusted to a standard distance of 1000-km so that range is not a factor in determining the phase function. Objects at large phase angles are backlit by the Sun so they usually appear fainter, while those at smaller angles are frontlit and appear brighter. There are exceptions to this behavior though. For example, Mallama et al (2025) determined that Starlink direct-to-cell satellites are bright at large and small phase angles but dimmer at intermediate angles.

BB satellites respond to phase angle in the usual manner as shown in Figure 4. The slope for BB 6 appears steeper than Block 1 though. As with satellite elevation the best fit to BB 6 magnitudes is fainter than that for Block 1 throughout the observed span of phase angles.

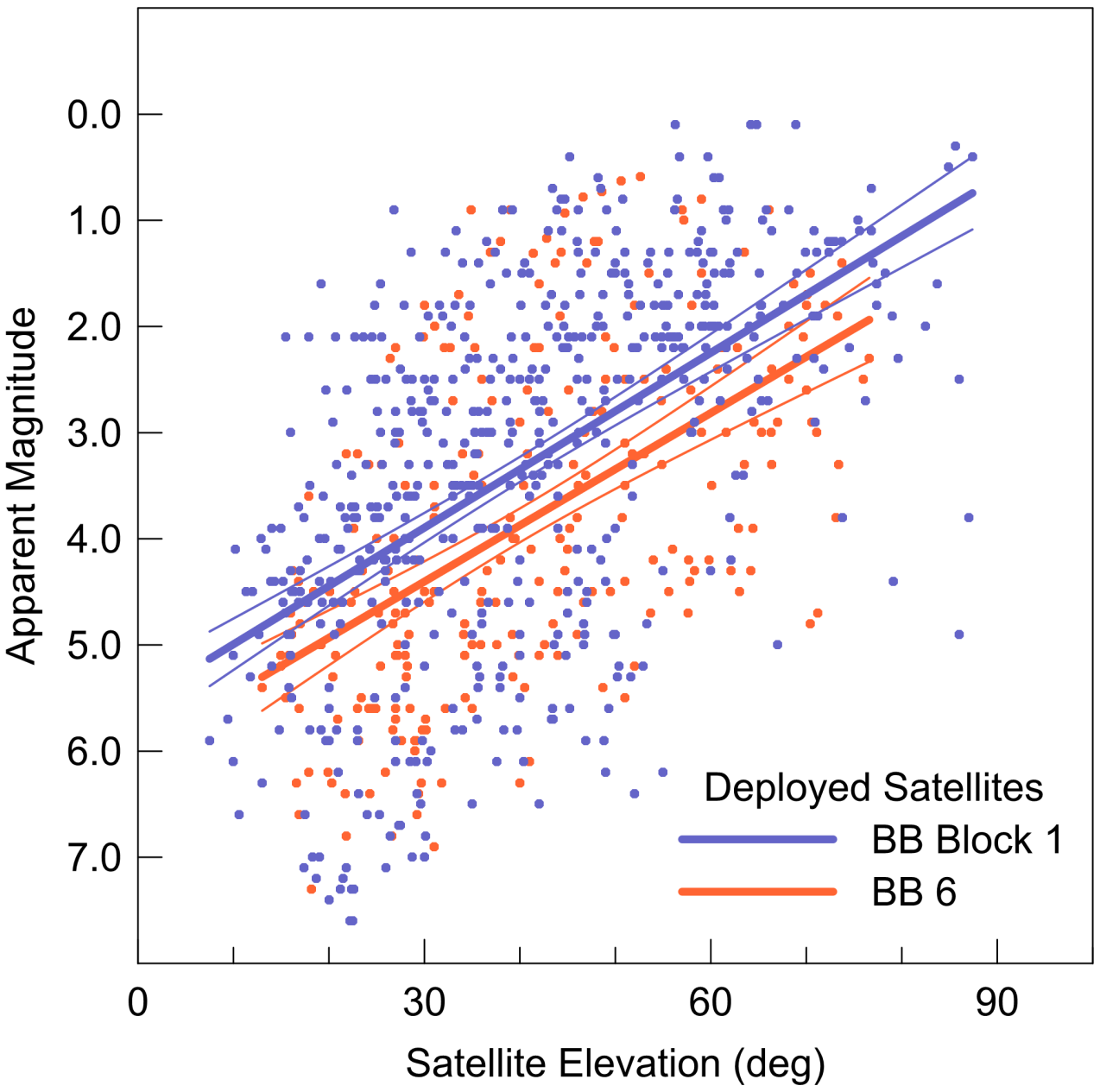

*Figure 3. The brightening of BB 6 and Block 1 satellites with elevation above the horizon. The trend line for BB 6 is fainter than that for Block 1 throughout the observed span of elevations.*

Finally, the beta angle is measured between a satellite's orbital plane and the direction to the Sun. Flat panel spacecraft are sometimes tilted towards the Sun when beta angles are high. This increases sunlight on their solar array which is mounted on the zenith facing side of the panel. Tilting reduces brightness because the nadir side facing observers on the ground is less illuminated. Apparent magnitudes are adjusted to 1000-km as with phase angle analysis so that range is not a factor in the beta function.

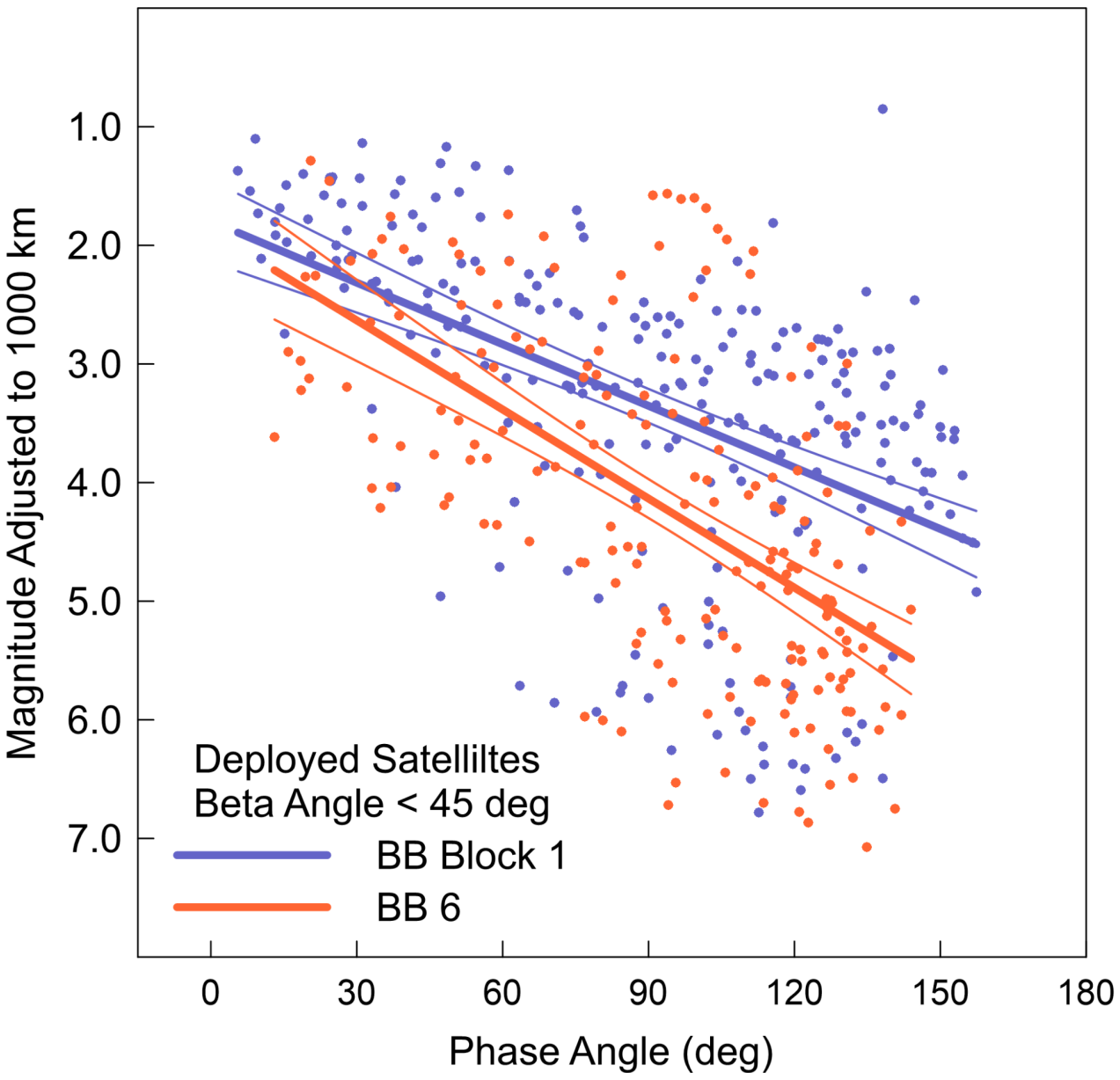

*Figure 4. Magnitudes adjusted to a uniform distance of 1000-km are plotted as a function of phase angle. The beta angles for magnitudes plotted in the graph are restricted to 45º and less to limit the effect of beta on brightness which is shown in Figure 5.*

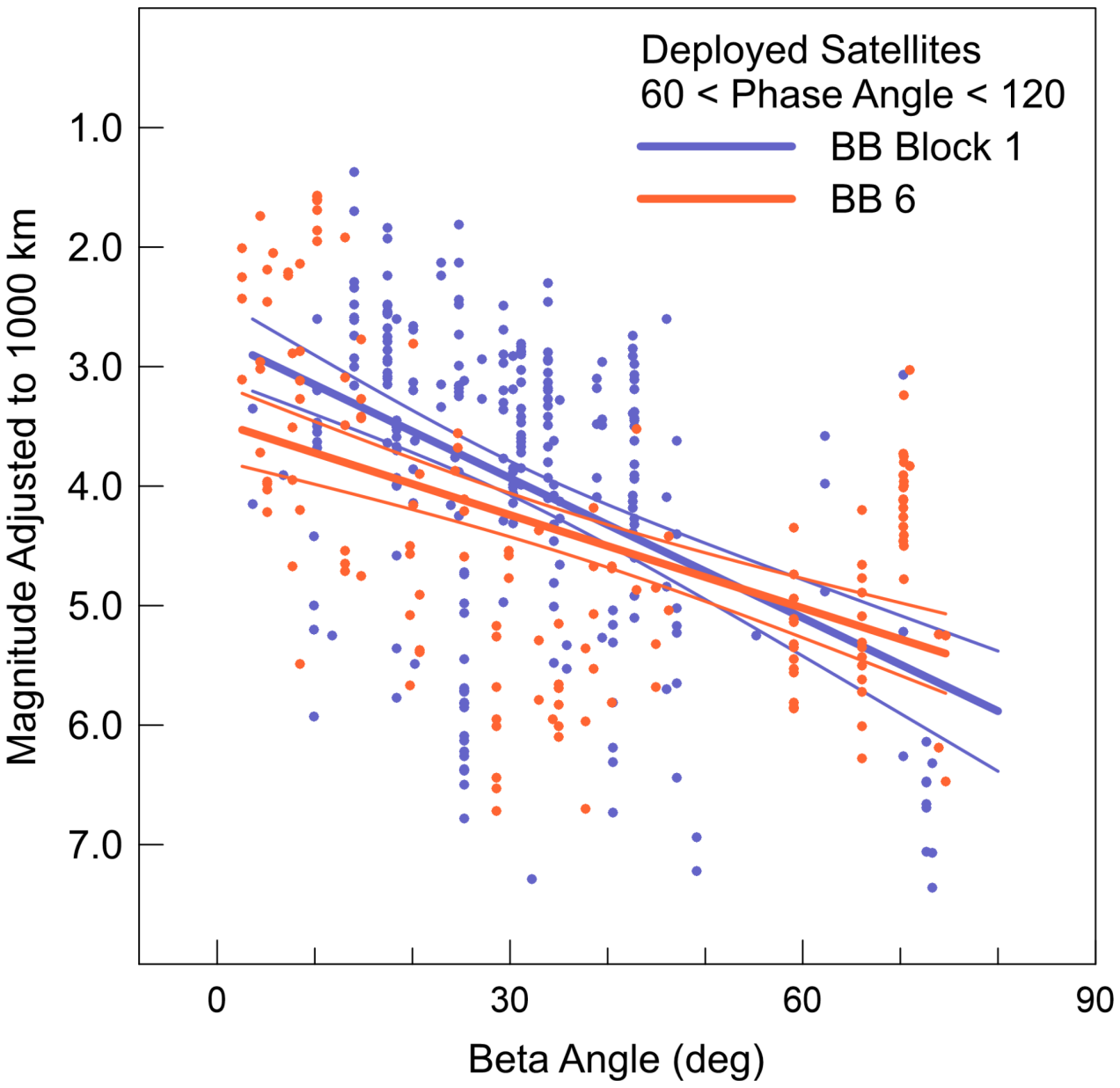

*Figure 5. Magnitudes adjusted to a uniform distance of 1000-km are plotted as a function of beta angle. The phases for magnitudes plotted in the graph are restricted to values of 90º +/- 30º to limit the effect of phase on brightness which is shown in Figure 4.*

Figure 5 shows that BB 6 and Block 1 spacecraft are fainter at large beta angles. This is consistent with tilting the solar array toward the Sun. The best fitting line indicates about 2 magnitudes of dimming over the span of observed angles. The slope may be larger for Block 1 satellites.

## 5. Brightness mitigation

AST SpaceMobile informed the US Federal Communications Commission (FCC) that “deployable antenna elements” are part of their brightness mitigation efforts for BlueBird (Cole et al 2025). These elements, shown in Figure 6, intercept sunlight before it reaches the Earth-facing side of the panel and they scatter it in a unique way.

The same paper modeled the impact of those structures on the luminosity of Block 1 spacecraft. The faintness of BB 6 relative to the smaller Block 1 suggests they are providing improved brightness mitigation. Resultant luminosity will depend on the spacecraft’s direction of motion relative to the deployable elements along with any tilting of the flat panel.

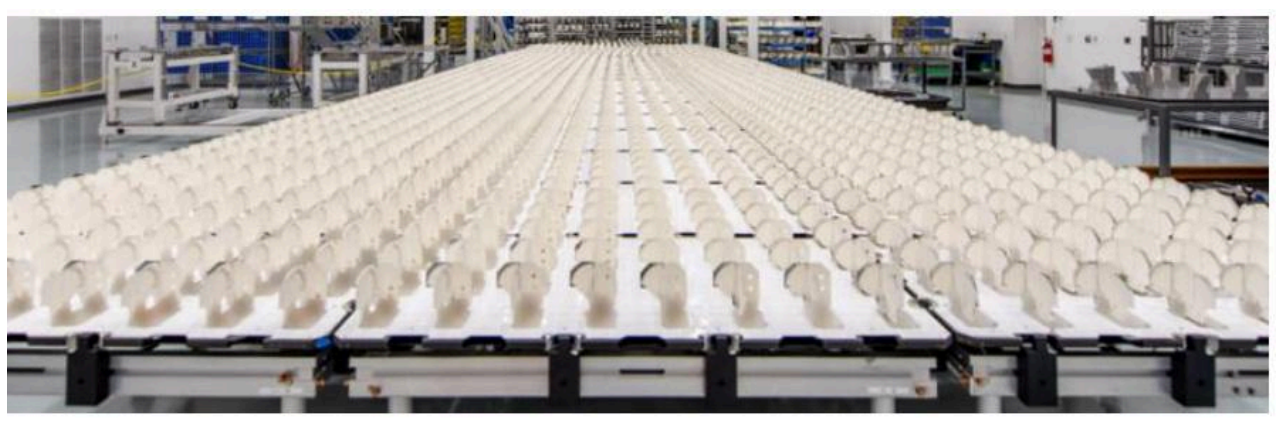

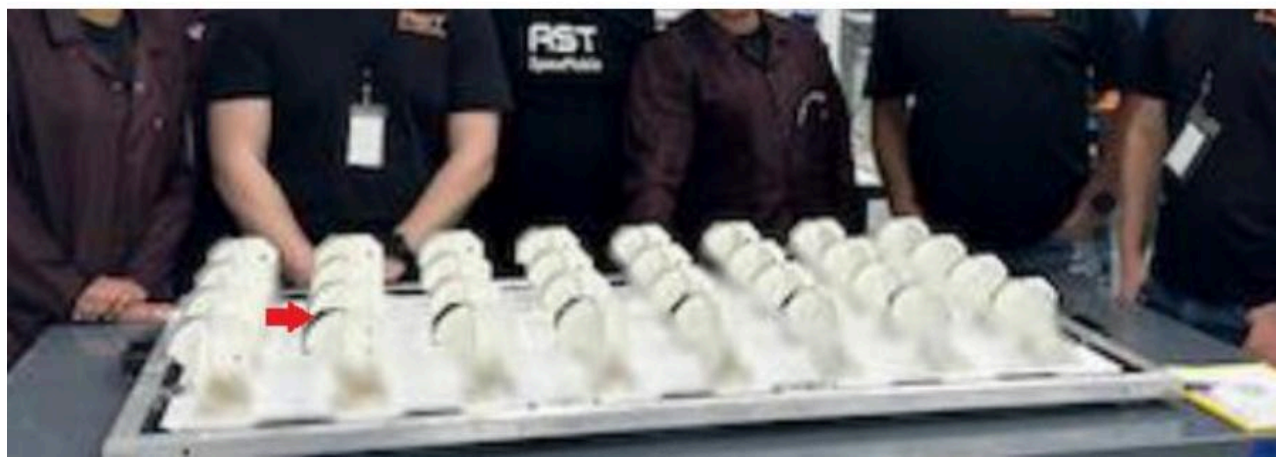

*Figure 6. Top: A wide angle view of an antenna panel. Bottom: A closeup of the deployable elements. (Credit: AST SpaceMobile)*

BB 6 was expected to be 3.48 times as bright as Block 1 satellites based on size scaling. That ratio equals 1.36 magnitudes, but BB 6 is actually 0.55 magnitude fainter as reported in Section 4. The 1.91 magnitude difference indicates a factor of 5.8 better brightness mitigation for BB 6 per unit of surface area. Put another way, BB 6 is only 17% as bright as Block 1 satellites in terms of area.

## 6. Impact on optical astronomy

BB satellites are a concern to astronomers performing observational research. The International Astronomical Union (IAU, 2024) recommended that brightness not exceed magnitude 7.0 for satellites at the altitude of BB. We refer to this as the *research limit*. The mean brightness of BB 6 exceeds this limit by 3.15 magnitudes which is a factor of 18.

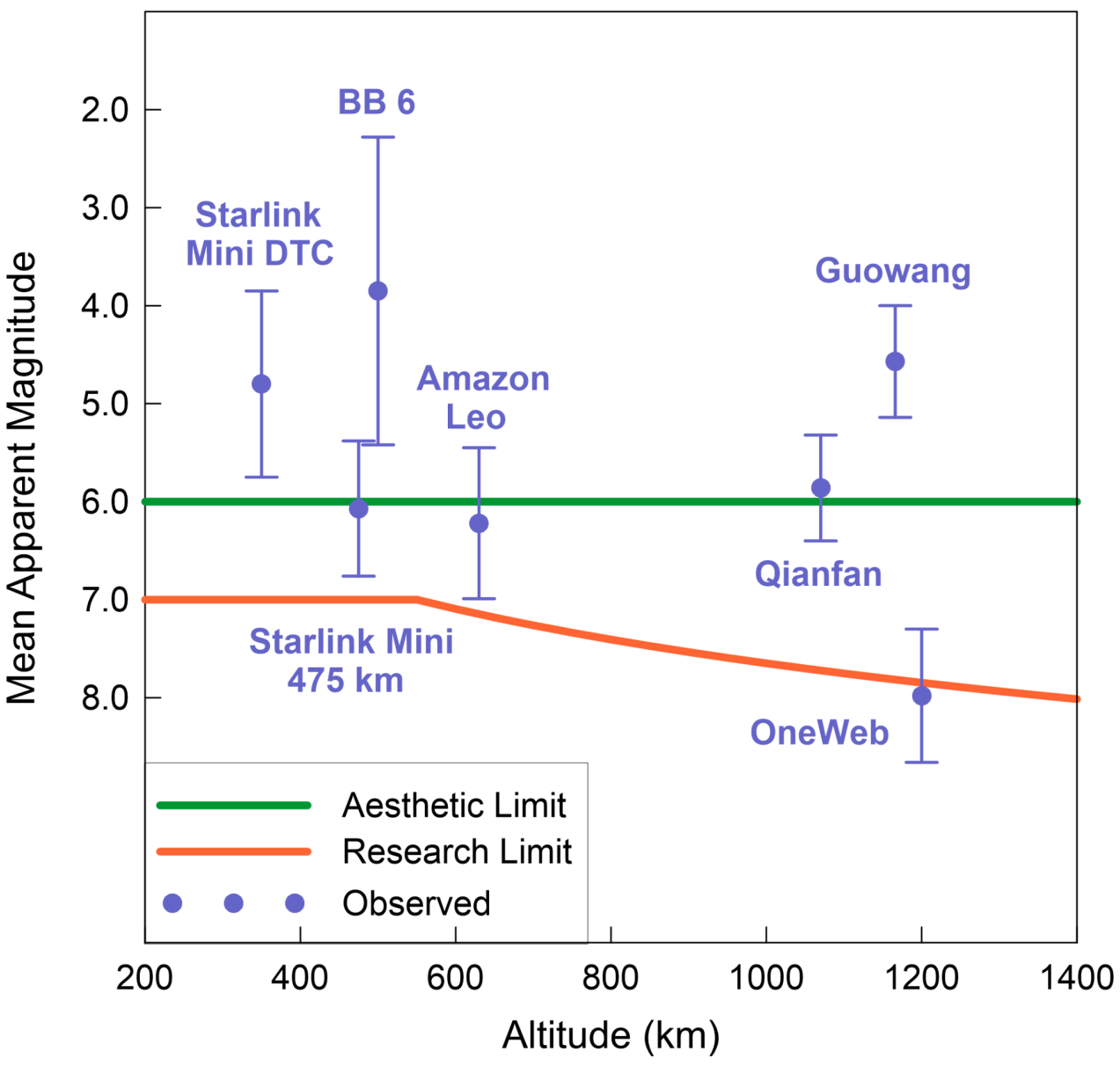

*Figure 7. The mean apparent brightness for BB 6 and its standard deviation are plotted at their 500 km altitude. The brightness of other constellation satellites along with the IAU recommended limits for non-interference with research and aesthetic appreciation of the sky are also shown.*

The IAU also stated that satellites in operational orbits should not be visible to the unaided eye. Objects of visual magnitude 6.0 can be seen at locations where the sky is minimally affected by light pollution. BB 6 exceeds this *aesthetic limit* by 2.15 magnitudes which is a factor of 7.

BB 6, the first satellite of Block 2, is more luminous than any large satellite constellation currently being launched, as shown in Figure 7. An offsetting factor, though, is that the BB satellites are expected to be fewer in number than several other constellations. The FCC has given AST SpaceMobile permission to launch just 248 BB satellites. By comparison there are already more than 10,000 Starlink spacecraft in orbit with plans for thousands more. The Qianfan and Guowang satellites are also expected to exceed 10,000 in number. Even greater numbers of AI spacecraft may be launched in the next few years.

Nevertheless, Hainaut (2026) reported that, “Bright satellites, such those from AST SpaceMobile, significantly impact saturating detectors even when their number is moderate.”

Magnitude statistics for all constellations were published by Mallama and Cole (2025). Updates to those results including new constellations and new models of satellites are provided on our website[4]. The means and standard deviations may be used to assess the impact of BB and any other constellation on optical astronomy.

## 7. Conclusions

The mean apparent magnitude for BB 6 is 3.85 +/- 0.10 while that for Block 1 spacecraft is 3.30 +/- 0.07. So, BB 6 is 0.55 magnitudes fainter. Likewise, when the apparent magnitudes are adjusted to a uniform distance of 1,000 km, the means are 4.32 +/- 0.08 and 3.77 +/- 0.06, respectively, and the difference is also 0.55 magnitude.

The faintness of BB 6 relative to Block 1 was unexpected because the BB 6 antenna panel is ~3.5 times as large as Block 1. In fact, the BB 6 spacecraft is only 17% as bright as Block 1 per unit surface area. While BB satellites exceed the brightness limits recommended by the IAU they are fewer in number than other constellations.

## Data availability

The observations studied in this article are available in the SCORE archive[5] and in the Mini-MegaTORTORA (MMT9) online database[6]. Software used for the analysis is available from the corresponding author.

## Acknowledgements

M. Dickinson of the IAU-CPS conducted a review of this study. Suggestions from two anonymous reviewers helped to improve the paper.

We thank S. Karpov for his correspondence about MMT9 and E. Katkova for maintaining the public database of observations.

[4] https://satmags.netlify.app/

[5] https://score.cps.iau.org/

[6] http://mmt9.ru/satellites/